\documentclass[pre,aps,twocolumn,floatfix,superscriptaddress,showpacs]{revtex4}
\usepackage{amsbsy}
\usepackage{amssymb}
\usepackage[dvips]{graphicx}

\begin{document}
\title{Energy exponents and corrections to scaling in Ising spin glasses}

\author{J.-P. Bouchaud}
\affiliation{
Service de Physique de l'\'Etat Condens\'e,
Orme des Merisiers --- CEA Saclay, 91191 Gif sur Yvette Cedex, France.}

\author{F.~Krzakala}
\affiliation{Laboratoire de Physique Th\'eorique et Mod\`eles Statistiques,
b\^at. 100, Universit\'e Paris-Sud, F--91405 Orsay, France.}
\affiliation{Dipartimento di Fisica, INFM, SMC,
Universit\`a di Roma La Sapienza P. A. Moro 2, 00185 Rome, Italy.}

\author{O.~C.~Martin}
\affiliation{
Service de Physique de l'\'Etat Condens\'e,
Orme des Merisiers --- CEA Saclay, 91191 Gif sur Yvette Cedex, France.}
\affiliation{Laboratoire de Physique Th\'eorique et Mod\`eles Statistiques,
b\^at. 100, Universit\'e Paris-Sud, F--91405 Orsay, France.}

\date{\today}


\begin{abstract}
We study the probability distribution $P(E)$ of the ground state energy $E$
in various Ising spin glasses. In most models,
$P(E)$ seems to become Gaussian with a variance growing
as the system's volume $V$. Exceptions include the
Sherrington-Kirkpatrick model (where the variance
grows more slowly, perhaps as the square root of the volume), and mean field
diluted spin glasses having $\pm J$ couplings.
We also find that
the corrections to the extensive part of the disorder averaged energy grow as a
power of the system size; for finite 
dimensional lattices, this
exponent is equal, within numerical
precision, to the domain-wall exponent $\theta_{DW}$.
We also show how a systematic expansion of $\theta_{DW}$ 
in powers of $e^{-d}$ can be obtained for Migdal-Kadanoff lattices. 
Some physical arguments are given to rationalize our findings. 
\end{abstract}
\pacs{PACS Numbers~: 75.50.Lk, 05.50.+q}
\maketitle

\section{Introduction}
\label{sect:intro}

Exponents and corrections to scaling play a central role in disordered
systems. Consider for instance the dependence of thermodynamic quantities
on the size $L$ of a system when $L \to \infty$.
In paramagnetic systems, correlations
are short range and thus bulk properties converge rapidly
to their thermodynamic limits. On the contrary, in
systems such as spin glasses below $T_c$, correlations
are long range and so finite size corrections are ``large''. In
practice, that means that even if one takes periodic
boundary conditions, the disorder average of an intensive quantity converges
to its thermodynamic limit slowly: finite size corrections go to zero as an
inverse power of $L$. This slow convergence is to
be contrasted with paramagnetic systems where finite size 
corrections in disorder averages are exponentially 
small. One can also consider 
the sample-to-sample fluctuations of thermodynamic quantities. These
fluctuations scale in the thermodynamic limit;
an interesting open question is 
their limiting distribution, if any.

In this work we consider these issues in the context of
Ising spin glasses~\cite{MezardParisi87b,Young98} at zero 
temperature and focus on the ground state energy.
Given a distribution of disorder realisations (each being represented
by $J$), the extensive ground state energy
$E_J$ is a random variable.
Denote by $N$ the number of spins in the system; 
$N=L^d$ for a $d$-dimensional hyper-cubic lattice.
We are interested in the probability distribution $P_N(E_J)$ and
in how the connected moments (cumulants) 
of this distribution
depend on $N$ (or $L$). Denoting disorder averages by an overline,
we have
\begin{equation}
\label{eq:Theta_s}
\overline {E_J(L)} = e_0 L^d + e_1 L^{\Theta_s} + \cdots
\end{equation}
In this expression, $\Theta_s$ is what we call the shift exponent.
To leading order, the energy scales with the volume,
while $\Theta_s$ gives the leading {\em correction} to scaling.
The justification for this notation will become clear later;
in the mean time, we should
compare to the usual notation involving
the correction to scaling exponent $\omega$; 
indeed, in the spin glass phase, 
finite size corrections of {\em intensive} quantities 
decay as a power of $L$:
\begin{equation}
\label{eq:omega}
\overline {E_J(L)/L^d} = e_0 + e_1 L^{-\omega} + \cdots
\end{equation}
and so $\omega = d-\Theta_s$.

We will also look at the width of $P_N$ which gives the 
{\em scaling} of the instance to
instance fluctuations:
\begin{equation}
\label{eq:Theta_f}
\left[ ~ \overline {E_J^2(L)} - \overline {E_J(L)}^2 ~ \right]^{1/2} = 
\sigma_0 L^{\Theta_f} + \cdots 
\end{equation}
$\Theta_f$ being the fluctuation exponent. Naturally,
one may extend these types of expansions to any cumulant of
$P_N$, a central question being whether 
$P_N$ has a limiting shape as $N\to \infty$. One
motivation for this is the fact that the ground state energy
is an extreme statistic: there are 
$2^N$ configurations of spins ($S_i = \pm 1$) and
one is interested in the one of minimum energy.
(Note that these $2^N$ random energies
are correlated.) Little is known
about such statistics except in a few solvable 
cases~\cite{Galambos87,CarpentierLedoussal01}.
The purpose of this work is to determine $P_N(E_J)$ 
numerically for a variety of spin glass models. From our
measurements we extract $\Theta_s$, $\Theta_f$, etc...
and compare these results to theory.

The outline of this paper is as follows.
We first consider expectations arising from
analogies with other systems as well as
some known results. In section~\ref{sect:models} we
describe the different types
of spin glasses used in this study. Then 
we give in section~\ref{sect:Theta_f} our estimates of the
fluctuation exponent $\Theta_f$.  This is followed
by a study of the probability distribution of the ground state
energy in section~\ref{sect:PofE}; for most models 
we find that it becomes Gaussian 
in the large system size limit. Then in section~\ref{sect:Theta_s}
we present our results for the shift exponent
$\Theta_s$. For all of these studies, we compare
the $d$-dimensional case to the theoretical predictions.
In section~\ref{sect:pmJ} we show that 
these exponents sometimes depend on the distribution of
the disorder variables (the spin-spin couplings).
Finally in section~\ref{sect:discussion} we discuss
and conclude this work.
Some details of the analytic computations are given in
the appendix.

\section{Clues from theory}
\label{sect:theory}

What values should be expected for the two exponents
$\Theta_s$ and $\Theta_f$? Suppose we start with
the random energy model (REM)~\cite{Derrida81} as a 
guide. Its ground state
energy $E_J$ has a Gumbel distribution
in the large $N$ limit~\cite{BouchaudMezard97}
with a variance of $O(1)$. Also,
the disorder averaged ground state energy
grows as $e_0 N + O(\ln N)$ at large $N$. These properties
show that $\Theta_s=\Theta_f=0$ and thus in the REM
finite size corrections are ``small'',
though clearly much larger than in a paramagnetic system.

Another model that can guide us is
the directed polymer in a random medium (DPRM)~\cite{Halpin-HealyZhang86}.
The problem on a tree is very similar to the REM, although subtle differences
appear~\cite{CarpentierLedoussal01}. The two dimensional 
case (with one space and one ``time'' 
dimension) can also be solved in much detail. In these cases, perhaps 
surprisingly, a {\em single} exponent $\theta$ describes the
scaling of three {\em a priori} unrelated quantities:
(1) the corrections to scaling of the disorder
averaged ground state energy $E_J$;
(2) the size of sample-to-sample fluctuations of $E_J$;
(3) the typical excitation 
energy of the lowest ``system-size'' excitation, that
is an excitation that is macroscopically different from the ground
state. These properties lead to the remarkable relation
$\Theta_s=\Theta_f=\theta$. The DPRM is thus described by
a ``one-parameter'' scaling theory.

It is plausible that an 
extension of the DPRM scaling theory may
apply to Ising spin glasses. At the heart of such a theory,
initiated by McMillan~\cite{McMillan84},
developped by Bray and Moore~\cite{BrayMoore86}
and extended by Fisher and Huse~\cite{FisherHuse86},
is the exponent $\theta_{DW}$. This exponent is analogous
to the $\theta$ of the DPRM, and gives the scaling of domain wall 
energies, $E_{DW} \simeq \Upsilon L^{\theta_{DW}}$; numerical 
estimates give $\theta_{DW}\approx 0.20$ in $d=3$ and 
$\theta_{DW}\approx -0.28$ in $d=2$. 
(More generally, a number of other $\theta$ exponents 
have been introduced for 
spin glasses; they are all associated 
with {\em excitation} energies; this
is to be contrasted with our $\Theta$ exponents that are associated
with {\em ground state} energies.) If we follow
the correspondence with the DPRM, we expect that
$\Theta_s=\theta_{DW}$, justifying our use
of a ``theta'' notation for $\Theta_s$. Physically, this equality 
corresponds to the fact that these
systems are sensitive to boundary conditions; 
for some samples, these conditions are such that a ``domain wall'' must
be present in the ground state. However the analogy 
with the DPRM certainly
breaks down for $\Theta_f$: for any short range 
spin glass in dimension $d$, Wehr and Aizenman~\cite{WehrAizenman90}
proved that $\Theta_f=d/2$.
This shows that the REM and the DPRM are {\em not}
good guides for finite dimensional spin glasses.

To have more realistic theoretical predictions, and in particular
to preserve $\Theta_f=d/2$, it seems necessary to work with models
having a microscopic Hamiltonian defined over configurations
of $N$ spins. 
One approach is to use hierarchical (Migdal-Kadanoff) lattices;
there, analytical computations as well as powerful numerical
methods are possible. We shall also consider mean field
spin glasses where spins are coupled amongst one-another at random
so there is no geometry to speak of. The $P_N(E_J)$ 
in such models can be
referred to as the mean field prediction. Strangely enough, little
is know about these systems so we will have to determine
their behavior numerically. In some cases these models
lead to surprizes as we shall soon see.

\section{Models and methods}
\label{sect:models}

We focus on three families of Ising spin glass models so that
the effects of geometry and dimension are apparent.
The Hamiltonian for these models is
\begin{equation}
H = - \sum_{<ij>} J_{ij} S_i S_j
\end{equation}
where $S_i = \pm 1$ and the $J_{ij}$ are quenched
Gaussian random variables of zero mean and variance
$1$ (except in the case of the
Sherrington-Kirkpatrick model which we discuss later). The sum
$\sum_{<ij>}$ is over nearest neighbor spins on a given
graph having $N$ nodes; the different models we consider
vary simply by the nature of that graph. 
  
Our first family of models are of
the Edwards-Anderson (EA) type~\cite{EdwardsAnderson75}: the graphs are
square or cubic lattices of linear size 
$L$ (thus $N=L^d$ when the dimension is $d$), and the edges connect
nearest neighbors only; we take periodic boundary conditions
in all directions.

\begin{figure}
\includegraphics[width=8cm, height=3cm,angle=0]{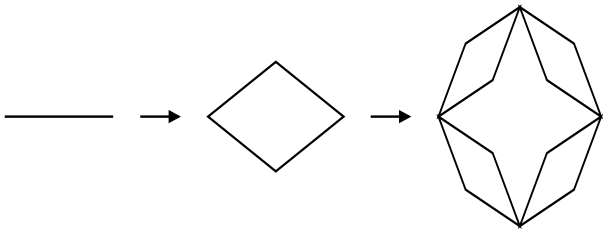}
\caption{Construction of a $d=2$ MK hierarchical lattice.}
\label{fig:MK}
\end{figure}

Our second family of models come from 
the Migdal-Kadanoff (MK) approach~\cite{SouthernYoung77}
where one performs
a bond-moving real-space renormalization group. This procedure
effectively amounts to computing quantities on
hierarchical (MK) lattices defined by an iteration process
(see Fig.~\ref{fig:MK}).
The iteration takes one bond (that is an edge of the
current graph) into $b$ paths, each
made of $2$ segments (that is edges); 
if $r$ is the iteration number (beginning with $r=0$),
the ``linear'' lattice size $L$
grows as $2^r$ and the volume (actually the number of edges
and thus the number of terms contributing to the energy) grows as
$(2 b)^r$.
When using $\ell$ segments instead of $2$ in each path, 
we have
\begin{equation}
L=\ell^{~r}  ~~ {\rm{and}} ~~ N\simeq(\ell ~ b)^r
\end{equation}
so that the dimension is $d=\ln N  / \ln L = 1 + \ln b / \ln \ell$.
The usual choice to obtain $d=3$ is $\ell=2$ and $b=4$, while
$\ell=b=2$ corresponds to $d=2$ as in Fig.~\ref{fig:MK}.

Our third family of models come from mean field, and here 
we have considered two types
of graphs. First, we use complete graphs where all
vertices are connected, corresponding to 
the Sherrington-Kirkpatrick (SK) model~\cite{SherringtonKirkpatrick75}.
To have an extensive energy, one takes the variance
of the $J_{ij}$ to be $1/N$.
Second, we also use diluted models for
which the connectivity is fixed and identical for
all the vertices of the graph~\cite{DominicisGoldschmidt89}.
The disorder ensemble then consists of the uniform distribution
over all graphs satisfying that constraint in addition
to the disorder ensemble in the $J_{ij}$.
Such an ensemble can be used to ``model'' the Euclidean case
by setting the coordination to that of the
lattice of interest. Thus to model the $d$-dimensional EA model
on the hypercubic lattice, we set the coordination to $2d$.

For these three families of spin glass models, we will determine
the distribution of the ground state energies. The
hierarchical lattices allow one to write a recursion for $P_N(E_J)$;
because of that, there is no need to {\em sample} the disorder variables, we
can perform the disorder average exactly.
On the contrary, for the EA and the mean field cases, 
we must compute the actual ground state energies for a large number
of disorder samples. For that we rely on a previously
tested~\cite{HoudayerMartin01} algorithmic procedure where
given enough computational ressources,
the ground state is obtained with a very high probability for both
Euclidean lattices and for random graphs as long as $N$ is
not too large. For our runs, we used several months
of CPU time on Pentium III personal computers 
running at 666 MHz. With this
amount of CPU, we obtained high statistics for
lattices of sizes up to $L=10$ (in both $d=2$ and $3$)
and for mean field graphs with
$N$ up to $300$.

\section{Sample to sample fluctuations and the
exponent $\Theta_f$}
\label{sect:Theta_f}

\subsection{Migdal-Kadanoff lattices}
\label{subsect:Theta_f_MK}
Let us begin with the Migdal-Kadanoff approach for 
which the important prediction 
$\Theta_f=d/2$ can be derived.
To understand how this relation comes about on the 
hierarchical lattices, we construct these by ``aggregation'',
i.e., by recursively (and hierarchically) joining sub-lattices together.
(This procedure is to be distinguished from the top-down iteration
used in Fig.~\ref{fig:MK}.) We work with
the distribution of ground state energies $E_p$ (respectively $E_a$)
subject to fixed boundary conditions, the spins
on the ends of the lattice being forced to be 
parallel (respectively antiparallel.) Let $\sigma^2$ be the
variance of the ground state energy
at some level of the hierarchical construction. To go to the next
level for which $L$ will be $\ell$ times larger,
first find the ground state energy in one
of the $b$ paths. Clearly, if we take the ground state configuration
in each of the segments of that path, we will have built the ground
state for {\em unconstrained} end spins. If the result does
not give the imposed values for the end spins
(parallel or antiparallel), one must add
a ``correction'' term equal to the smallest domain wall
energy of the $\ell$ segments of that path.
Thus the energy of one path is the sum of the 
ground state energies of the $\ell$ segments,
plus one domain wall energy half of the time.
Second, we add up the {\em independent} contributions
from the $b$ different paths, leading to 
$E_p$ and $E_a$ at the new level. The ground state energy 
is then simply $min(E_p,E_a)$. If we neglect the ``correction''
term, then $\sigma^2$ at this new
level is just $\ell b$ times larger than at the previous level.
At large $b$, the correction is in fact small and so
it can be neglected. From one level to the next, 
the volume grows by a factor $\ell b$, 
just as $\sigma^2$ does, so the variance is {\em linear} in the lattice
volume and thus $\Theta_f=d/2$. From our numerical study
of these hierarchical lattices, we find that this relation
holds also for small $b$ and for both $\ell=2$
and $\ell=3$. (We did not test for larger $\ell$.)

As mentioned in the introduction, the
Wehr-Aizenman theorem~\cite{WehrAizenman90}
shows that $\Theta_f=d/2$
in finite dimensional spin glasses. It is rather conforting
that the MK approach also leads to this result,
sustaining the belief that it is a useful guide for
real (finite dimensional) spin glasses.

\subsection{Mean field models}

Next, consider the mean field
prediction for $\Theta_f$. Since there is no geometry in our
mean field family of models,
we identify $L$ with $N^{1/d}$ and thus
\begin{equation}
\label{eq:Theta_f_MF}
\left[ ~ \overline {E_J^2(N)} - \overline{E_J(N)}^2 ~ \right]^{1/2} = 
\sigma_0 N^{\Theta_f/d} + \cdots 
\end{equation}
The issue here is whether mean field also predicts $\Theta_f/d=1/2$
as it should if one believes that this approach
correctly describes the large dimensional limit of 
real spin glasses. To find out, we have performed
ground state computations on thousands of samples of
the SK model and of fixed connectivity spin glasses,
and have extracted for each ensemble its associated
$\Theta_f/d$.

First, consider the fixed connectivity spin glasses with 
the connectivities, $z=3,4,6$ and $10$.
To estimate $\Theta_f/d$, we determine when
the root mean square deviation (RMS) of the ground state energy
divided by $N^{\Theta_f/d}$ becomes flat as a function of
$N$. (Our runs were performed for $50 \le N \le 300$.)
This leads to $\Theta_f/d \approx 0.5$ for $z=3$,
$\Theta_f/d \approx 0.49$ for $z=4$,
$\Theta_f/d \approx 0.48$ for $z=6$,
and $\Theta_f/d \approx 0.44$ for $z=10$.
The drift we observe in $\Theta_f/d$ is most probably
an artefact of our procedure, and simply corresponds
to the fact that the corrections to the scaling
in Eq.~\ref{eq:Theta_f_MF} are important in our data,
especially at large $z$.
To get better estimates of $\Theta_f/d$, we would need
to control these corrections to scaling but our data
are not sufficiently precise for that. Nevertheless, it seems
very plausible that $\Theta_f/d=0.5$
in all the fixed connectivity models.
In direct analogy to what was stated previously, we can also say that
this result sustains the belief that mean field models
provide a useful guide to real (finite dimensional)
spin glasses.

Let us now continue and consider 
the limit of infinite connectivity, i.e., the SK model.
Since there are $O(N^2)$ terms contributing to the
Hamiltonian, a very simple minded guess would give 
$\Theta_f/d = 1$ and thus
larger fluctuations than in the other models. However
the opposite happens, revealing that the scaling in
the SK is quite subtle. One can get some clues from
Kondor's analytic study~\cite{Kondor83} performed just
below the critical temperature $T_c$. In particular, his 
results have been interpreted 
by Crisanti et al.~\cite{CrisantiPaladin92} who argued that 
the {\em free energy} 
fluctuations should scale as $N^{1/6}$, which is very small 
compared to the na\"{\i}ve
estimate. A different estimate was recently proposed
by Aspelmeier et al.~\cite{AspelmeierMoore03b}
who argued that free energy fluctuations should
scale as $N^{1/4}$. Our concern here is the ground state energy;
although there are no analytical
calculations, it is plausible 
that the exponent for energy fluctuations at $T=0$ is 
the same as at that for free energy fluctuations at $0<T<T_c$.
(Such an extrapolation to $T=0$ is known to apply to the DPRM where 
the exponent of the free energy fluctuations 
at $T >0$ is equal to the one of the 
ground state energy fluctuations.)
In that case, we would have $\Theta_f/d=1/6$
according to some authors and $\Theta_f/d=1/4$ 
according to others.

\begin{figure}
\includegraphics[width=8cm, height=7cm,angle=0]{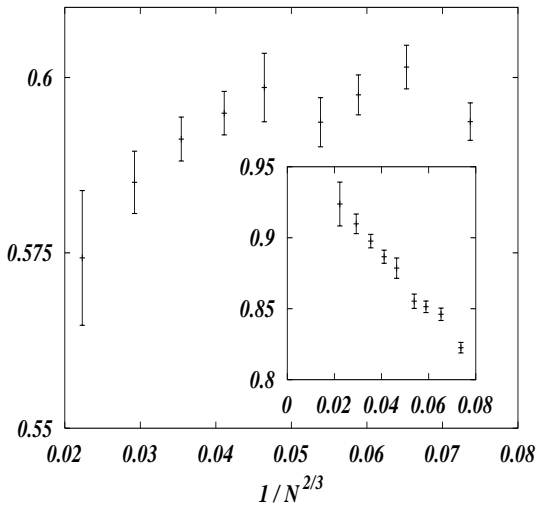}
\caption{Scaled root mean square deviation (RMS) of the ground state energy
in the SK model, displayed as a function of $N^{-2/3}$. 
Main figure: RMS divided by $N^{1/4}$; inset:
RMS divided by $N^{1/6}$.}
\label{fig:SK_sigma}
\end{figure}
What do the numerical estimates tell us about this question?
We are aware of a study by Cabasino et al.~\cite{CabasinoMarinari88}
who showed beyond any reasonable doubt that $\Theta_f/d < 0.5$;
in fact their best fit gives $\Theta_f/d \approx 0.28$. When 
we consider our data, motivated by Eq.~\ref{eq:Theta_f_MF}, 
we find that the ratio of the left and right hand sides
is compatible with a constant when we use $\Theta_f \approx 0.25$
and $40 \le N \le 150$, but the two points at $N=200$ and
$300$ are then below the others as show in 
Fig.~\ref{fig:SK_sigma}.
It is difficult to extract an error bar on the value of
this exponent, and most likely the terms dropped in 
Eq.~\ref{eq:Theta_f_MF} are important just as was
the case in the fixed connectivity models. In fact, here the 
difficulty is even more acute as 
the standard deviation increases only very slowly
with $N$. Nevertheless, let us fit the standard deviation
to a {\em pure} power; 
then using a $\chi^2$ analysis we obtain
$\Theta_f/d = 0.24 \pm 0.005$
which is very close to the conjectured $1/4$ value,
a result further supported by other numerical
work~\cite{PalassiniPrivate}.
Note however that throughout all this paper,
the error bars given are statistical only\ldots
In our fit, there are
$7$ degrees of freedom and the resulting 
$\chi^2\approx 9.5$ is not bad; nevertheless, 
a critical examination of the figure leads one to conclude that 
the actual uncertainty on
$\Theta_f/d$ is certainly much greater than $0.005$. Note that this 
estimate is a bit lower that that
of~\cite{CabasinoMarinari88} and is
many standard deviations away from
the conjecture $\Theta_f/d = 1/6$.
However, if we impose $\Theta_f/d = 1/6$, the data is reasonably
straight as a function of $N^{-2/3}$ as shown in the inset 
of Fig.~\ref{fig:SK_sigma}. Because of this ``good'' behavior, we
cannot rule out $\Theta_f/d = 1/6$.

Given that in the dilute spin glasses we expect $\Theta_f/d = 0.5$, 
how does one recover the SK case as $z \to \infty$? 
Begin by recalling that, in the SK model, the 
variance of the $J_{ij}$ is taken to be $1/N$ to ensure that
the ground state energy is extensive. To maintain this property in
the diluted spin glasses, we must divide the
$J_{ij}$ by $\sqrt{z}$. The fluctuations for large $N$ and $z$ 
in these modified spin glasses scale as
\begin{equation}
\label{eq:Theta_f_MF_diluted}
\left[ ~ \overline {E_J^2(N)} - \overline{E_J(N)}^2 ~ \right]^{1/2} = 
\sigma_0(z) N^{\Theta_f/d} / \sqrt{z} + \cdots 
\end{equation}
with $\Theta_f/d=1/2$ from what we saw previously.
The SK limit corresponds to setting $z=N$, so we see
that smooth large $N$ and $z$ limits
require $\sigma_0(z) \simeq z^{\mu}$ where $\mu$
is equal to the SK value of $\Theta_f/d$.
We saw that it was difficult to obtain
$\Theta_f/d$ in the fixed connectivity models, but
obtaining the prefactor $\sigma_0(z)$ is even more difficult.
Nevertheless, we have extrapolated our data
for the root mean square deviation divided by $N^{0.5}$
in the different finite connectivity models. This
leads to the estimates
$\sigma_0(z=3) \approx 0.67$, 
$\sigma_0(z=4) \approx 0.70$, 
$\sigma_0(z=6) \approx 0.73$, 
and $\sigma_0(z=10) \approx 0.76$.
This growth is very slow; it is compatible with the value $\mu = 1/6$ 
but much less with the value $\mu = 0.25$. Given the large
uncertainties in our extrapolations however, this has to 
be considered as only very indirect evidence in favor
of $\Theta_f/d = 1/6$.

We conclude that understanding the non-trivial 
size dependence of the fluctuations of the ground 
state energy in the SK model remains quite a challenge. Let us propose here 
a simple argument suggesting that 
$\Theta_f/d = 1/4$. Imagine that one changes 
slowly all the couplings 
$J_{ij}=J_{ij}^0/\sqrt{N}$: $J_{ij}^0 \to J_{ij}^0 + \delta  J_{ij}^0$, 
where the order of magnitude of $\delta J_{ij}^0$ is denoted by $\varepsilon$. 
If $\varepsilon$ is infinitesimal, the ground state remains the same. Up 
to what value of $\varepsilon$ will this be true? The change of local field 
induced by the change of couplings for a fixed configuration $\{S_i^*\}$ 
of the spins is of order 
$\sum_j \delta  J_{ij}^0 S_j^* /\sqrt{N} \sim \varepsilon$.
Using the fact that the local field distribution vanishes linearly for small
fields in the SK model, it is easy to show that the smallest local field is
of $O(1/\sqrt{N})$. Therefore, the first value of $\varepsilon$ that will 
trigger a change of the ground state is $\varepsilon^* \sim 1/\sqrt{N}$. 
Flipping the spin with this low local field should lead to a 
cascade of flips that lowers the energy and thus to a new ground
state that is ``substantially'' different from the starting one.
Furthermore, the variation in the ground state energy when going
from $\varepsilon=0$ to $\varepsilon=\varepsilon^*$ is probably
$O(1)$; indeed,
there are excited states of energy $O(1)$ above the ground 
state, and one 
therefore expects level crossings to occur
when $\varepsilon \sim 1/\sqrt{N}$. Finally,
in order to scan
the whole distribution of $J_{ij}^0$'s, one needs $\varepsilon \sim 1$. 
Using this range, the ground state will 
change $1/\varepsilon^* \sim N^{0.5}$
times. Since between each level crossing, the ground state 
energy randomly changes by an 
amount $O(1)$, the total expected fluctuation of the 
ground state energy will be 
$O(N^{1/4})$, so that $\Theta_f/d = 1/4$.
More analytical work is obviously needed to confirm this 
speculative result, but note
that at the heart of our argument lies the 
fact that the ground state of the SK model
is particularly {\em fragile}: a relative change of 
order $1/\sqrt{N}$ of the disorder 
is enough to substantially change the ground state.

\subsection{Edwards-Anderson models}

\begin{figure}
\includegraphics[width=8cm, height=7cm,angle=0]{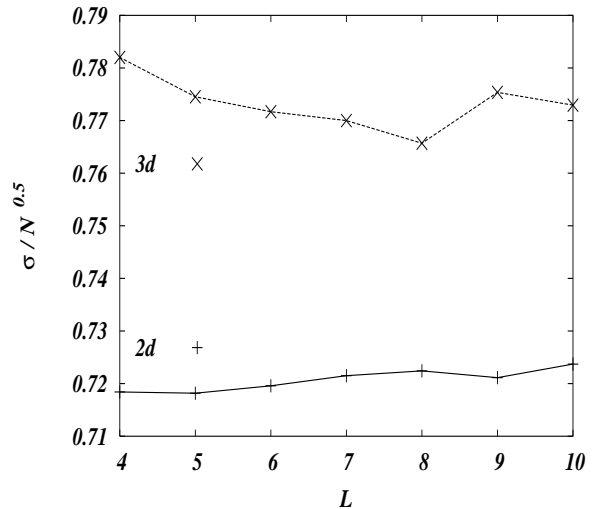}
\caption{Standard deviation of the ground state energy
divided by the square root of the volume for the EA model
in $d=2$ and $3$, Gaussian $J_{ij}$s.}
\label{fig:sigmaRescaledEA}
\end{figure}

In the case of the finite dimensional lattices, we know
that $\Theta_f=d/2$ holds exactly because of
the Wehr-Aizenman theorem~\cite{WehrAizenman90}, and recently
Aspelmeier and Moore~\cite{AspelmeierMoore03a} found
that this relation holds within replica
theory. In spite of these
theoretical results, it is instructive
to see how this equality transpires numerically.
We thus follow the procedure used in the fixed 
connectivity models where we tested for when
the rescaled RMS became size independent. In $d=3$, the
rescaled data show no obvious trend 
when $1.49 \le \Theta_f \le 1.60$ 
while in $d=2$ the corresponding range is
$1.00 \le \Theta_f \le 1.02$. In particular,
in Fig.~\ref{fig:sigmaRescaledEA} we
show these ratios when $\Theta_f$ is 
set to $d/2$ (the data displayed are from the models 
with Gaussian $J_{ij}$s.)

In principle, it would be interesting to find
the {\em corrections} to this scaling law. In a
renormalization group picture, corrections go
as inverse powers of $L$. Furthermore, if one believes 
that the exponent $\Theta_s$ gives the leading corrections
to scaling for {\em all} extensive quantities, then 
those for the variance of $E_J$ 
should go as $L^{\Theta_s}$. To consider this
possibility, we set $\Theta_f=d/2$ and then ask when 
our data for
the rescaled RMS follow approximately
a straight line when plotted as a function of $L^{-\omega}$
with the expectation that $\omega = d-\Theta_s$.
In $d=3$, we find $\omega \geq 1.5$, to be compared with
the ``theoretical'' value $\omega \approx 2.8$ (cf.
section~\ref{sect:Theta_s}). Since
$\Theta_s$ is small and may in fact be zero, we have
performed the analysis for $\omega=3$; using the intercept
of the curve with the $y$ axis, we find
that the RMS of $E_J$ grows 
as $0.765 L^{3/2}$.
In $d=2$, the rescaled data is too flat and so in practice we
cannot give any sensible estimate of $\omega$. But we can 
follow the $d=3$ procedure, setting this time $\omega=2$
($\Theta_s$ is also relatively close to zero in $d=2$);
the corresponding fits give that
that the RMS of $E_J$ grows as
$0.725 L$. The moral of this story is that
corrections to scaling are in general very difficult to
determine, even if the leading scaling 
law is known exactly. Nevertheless, it seems that
the same $\omega$ may very well describe the
dominant corrections to scaling of many observables,
as expected from the renormalization group picture.

To put these last numbers in perspective, consider
the Mattis model where the couplings are gauge transformed
from a ferromagnet having $J_{ij}$ chosen randomly
on the positive side of a Gaussian. The total ground
state energy of such a system is the sum of all these couplings
and thus has the expectation value $dN \sqrt{2/\pi}$ on the
hypercubic lattice of dimension $d$. On the other hand its 
variance is $d N [1 - 2/\pi ]$. Thus in this Mattis model, 
the RMS of the ground state energy grows as 
$1.04 L^{3/2}$ in $d=3$, and as
$0.852 L$ in $d=2$. As expected, the Mattis model has
larger absolute fluctuations than the EA model. It
is also appropriate to compare
the {\em relative} fluctuations $\sigma_r$, 
that is the RMS of the ground state energy divided by its
mean. For the Mattis model, we find $\sigma_r = 0.434 / L^{3/2}$
in $d=3$ and $\sigma_r = 0.534/L$ in $d=2$. These should be
compared to the values we find in the EA model:
$\sigma_r = 0.450 / L^{3/2}$ in $d=3$ and 
$\sigma_r = 0.551 / L$ in $d=2$. This shows that the
relative fluctuations are 
slightly {\em smaller} in the Mattis model than in
the EA model. Although this is in line with what 
frustration should do,
note that the size
of the effect is about $3\%$ which is very very small.

To summarize our study of $\Theta_f$ for the different 
spin glass models, we have found that
all the models considered seem to satisfy $\Theta_f=d/2$.
The notable exception is the SK model for which we made the
case that $\Theta_f=d/4$ was likely but
$\Theta_f=d/6$ was also possible; no matter what, 
$\Theta_f=d/2$ is excluded, showing
that the $N \to \infty$ and $d \to \infty$
limits do not commute. In all other cases, 
the variance of the ground state energy
grows linearly with the system's volume; this is the scaling 
expected from a central limit theorem behavior
when the different terms contributing
to the ground state energy are independent. Our results thus tell us 
that these terms are only {\em weakly} correlated.
This feature is completely missed by both the REM and the DPRM; 
although a one-parameter scaling picture applies to those two models,
it cannot apply to spin glasses. Note that this central limit
behavior suggests that $P_N(E_J)$ tends to a Gaussian; we now
turn to see whether this is the case.

\section{Probability distribution of ground state energies}
\label{sect:PofE}

If the central limit theorem (CTL) were
applicable, not only would we have $\Theta_f=d/2$ for the 
scaling of the fluctuations, but also
the shape of $P_N(E_J)$ 
would become Gaussian at large $L$. This behavior indeed arises
for the MK lattices, both
analytically at large $b$ and numerically for all $b$.
(In our numerical study, we find that the skewness
and kurtosis $P_N(E_J)$ decrease fast towards zero as
$L$ grows.) Obviously,
the terms contributing to the ground state energy are
not independent but their correlations are not strong
enough to prevent a CLT large $L$ scaling. 

\begin{figure}
\includegraphics[width=8cm, height=7cm,angle=0]{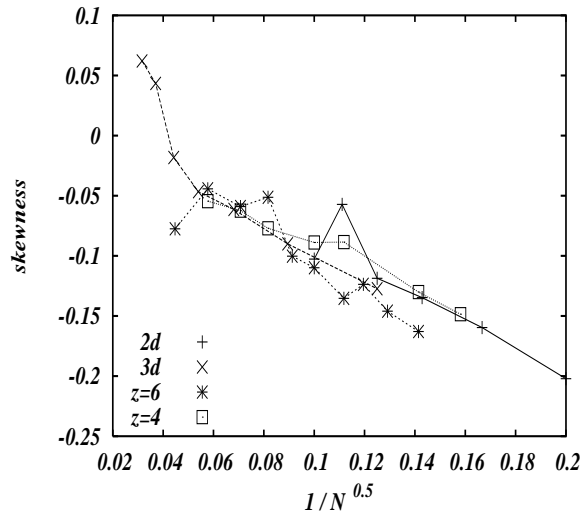}
\caption{Skewness of the ground state energy distribution
for mean field and EA models with Gaussian $J_{ij}$s.}
\label{fig:skewness}
\end{figure}

The question we address here is whether this simple behavior also holds
in the other models. Let us begin with the mean field case.
For the fixed connectivity mean field graphs, our data for
the skewness and kurtosis of $P_N(E_J)$ decrease in magnitude
as $N$ increases; this decrease is 
compatible with an extrapolation to zero as $N \to \infty$
as illustrated in Fig.~\ref{fig:skewness}
(These quantities are difficult to measure to high
precision, so this should be considered
as only suggestive of a Gaussian limit for $P_N(E_J)$.)
The SK model however is clearly in a different class:
not only does it have $\Theta_f \ne d/2$ but also its 
$P_N(E_J)$ is {\em not} Gaussian. In particular, its
skewness at large $N$ stabilizes around
$-0.43 \pm 0.02$, while its kurtosis stabilizes around $0.40 \pm 0.03$.
It is instructive to compare this to the values 
predicted by the REM model
where $P(X_J)$ is a Gumbel distribution: there, the
skewness is $-1.139$ while the kurtosis is exactly $2.4$.
Our estimates do not agree either with
the values from the Fisher or Weibull
universality classes:
we thus conclude that the
SK ground state energy distribution does not belong to one of the known
universality classes of extreme statistics~\cite{BouchaudMezard97}.

Finally we come to the EA lattices. Given that the MK and 
diluted mean-field graphs lead to the same
conclusion, it will come as no surprise that
our data for the EA lattices are also compatible
with a Gaussian limiting shape
for $P_N(E_J)$.
Note that this is expected though not proven from
the work of Wehr and Aizenman~\cite{WehrAizenman90},
while it does follow from 
replica theory calculations~\cite{AspelmeierMoore03a}.
An examination of the skewness and kurtosis of
$P_N(E_J)$ shows that they decay with system size,
both in $d=2$ and $d=3$. Although our measurements
lack precision when the number of spins is large, 
the extrapolations
suggest zero limiting values as $L \to \infty$
as one can see in Fig.~\ref{fig:skewness}. 
Perhaps it is also worth noting
that our data are {\em quantitatively} similar in
the different models; for instance, for $L=8$ and $d=3$,
we find the skewness to be $-0.18$ in EA and $-0.21$ in MK.

The overall situation indicates that 
$P_N(E_J)$ is Gaussian. Furthermore we checked whether the
convergence to this
Gaussian follows the central limit
theorem law. Indeed, that law predicts for instance that
the skewness scales as $N^{-1/2}$; thus
we have plotted in Fig.~\ref{fig:skewness}
the skewness for the different models as a function of that
scale. The data is completely compatible with a {\em linear}
convergence to zero, confirming the CLT scaling.
Only in the case of the SK model are the variables
contributing to the ground state energy sufficiently correlated to 
prevent a central limit theorem behavior. A physical
interpretation of this is that as soon as there is
a kind of {\em locality} which allows one
to decompose the sample into quasi-independent
subsystems, the central limit theorem behavior
will appear, leading to
$\Theta_f=d/2$ and a Gaussian $P_N(E_J)$.
Of course, it is not clear why this should apply
to the diluted mean field graphs.

\section{The shift exponent $\Theta_s$}
\label{sect:Theta_s}

\subsection{Migdal-Kadanoff lattices}

We now move on and study the
finite size corrections to the mean energy density. Following 
Eq.~\ref{eq:Theta_s}, the mean excess 
of the (extensive) 
ground state energy is expected to
scale as $L^{\Theta_s}$.
To have an idea of what this exponent should be, it is again most useful
to begin with the hierarchical lattices. The important 
prediction of that approach is that $\Theta_s=\theta_{DW}$
where $\theta_{DW}$ is the domain wall exponent.
To see why this is so, reconsider 
the evolution equation for the energies $E_p$ and $E_a$ as 
one applies the recursion. First,
along a given path, the energy is the sum of
the ground state energies of each of its $\ell$ segments,
the sum being sometimes corrected by
the energy of the domain wall of the weakest segment
in order to satisfy the boundary conditions. This 
correction shifts the path's
energy by $O(L^{\theta_{DW}})$. Second, adding
the energies of the different paths does not change the scale
of the shifts, so necessarily $\Theta_s=\theta_{DW}$.
Naturally, we have confirmed this relation numerically for different values
of $\ell$ and $b$ by
performing fits to moments of $P(E_p,E_a)$
(determined with no statistical error for these hierarchical lattices)
from which we extracted estimates for $\Theta_s$ and $\theta_{DW}$.

When $b$ is large, we can derive the analytical 
expression for $\theta_{DW}$ on 
these MK lattices. Indeed, 
in this limit, $P(E_p,E_a)$ becomes Gaussian and so one
can perform a cumulant expansion about this Gaussian, leading 
to a $1/b$ series. This scheme extends
the work of Southern and Young~\cite{SouthernYoung77} who assumed 
that $P(E_p,E_a)$ was Gaussian even for finite $b$. For general $\ell$ and 
$b$ we obtain 
\begin{equation}
\label{eq:MKresult}
\theta_{DW}(\ell,b) = {\frac{d}{2}} + a_0(\ell) + \frac{a_1(\ell)}{b} + 
O({\frac{1}{b^2}}) \ .
\end{equation} 
When $\ell=2$, we find $a_0=-1.2302$ and $a_1=-0.04573$ (see the
appendix for a derivation). In Fig.~\ref{fig:MK_theta_DW}
we plot the difference
between the numerically obtained $\theta_{DW}(\ell=2,b)$ and 
the terms of the expansion given in Eq.~\ref{eq:MKresult}. This
allows us to determine
numerically the next term of the expansion,
and we find $-0.045 \pm 0.001/b^2$.
This value could be obtained analytically, but we have not pushed
the analytical calculation to that order. 
\begin{figure}
\includegraphics[width=8cm, height=7cm,angle=0]{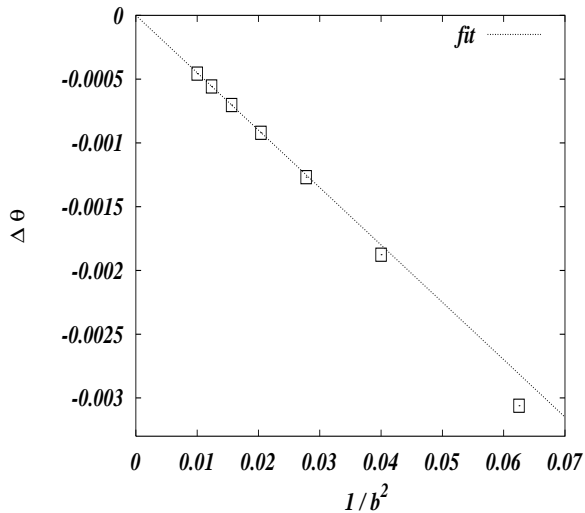}
\caption{Difference between 
$\theta_{DW}$ and the first three terms in the series
of Eq.~\ref{eq:MKresult} as a function of
$1/b^2$ where $b$ is the parameter of
the Migdal-Kadanoff lattices. The line is $-0.045 / b^2$.}
\label{fig:MK_theta_DW}
\end{figure}
Note that the $1/b$ expansion corresponds to an 
expansion in $e^{-d}$ where $d$ is the dimension of space.
This justifies the fact that the $1/b$ expansion is quite
accurate all the way down to $d=3$ (which corresponds to $b=4$).
Finally, when $\ell \to \infty$, we obtain $a_0=3/2$. These results
show that $\Theta_s = \theta_{DW} < \Theta_f$, 
justifying the neglect of the ``correction'' 
terms in section~\ref{subsect:Theta_f_MK} 
from which we concluded that $\Theta_f=d/2$.

Before going on to the mean field case, let us remark
that the MK value for $\theta_{DW}$ is quite close
the actual value in the EA model.
If we use the standard choice for $d=3$, 
$\ell=2$ with $b=4$, the MK prediction is
$\theta_{DW} \approx 0.255$. One can also use the choice
$\ell=3$ with $b=9$ for which $\theta_{DW}\approx 0.242$. These
values are to be compared to current estimates for $\theta_{DW}$ in
the EA model, $\theta_{DW} = 0.21 \pm 0.02$~\cite{PalassiniYoung99b}
and $\theta_{DW} = 0.19 \pm 0.02$~\cite{Hartmann99}.
A similarly good comparison occurs when $d=2$.

\subsection{Mean-field models}

We have seen that the prediction of the MK approach 
is $\Theta_s=\theta_{DW}$; 
in $d=3$, this gives either $\Theta_s = 0.25$ (the MK value)
or $\Theta_s \approx 0.20$ (if one uses the $d=3$ EA model
values for $\theta_{DW}$). In the case of 
the mean field models, there is no way to introduce domain
walls, and so we simply focus on their prediction for
$\Theta_s$. 

\begin{figure}
\includegraphics[width=8cm, height=7cm,angle=0]{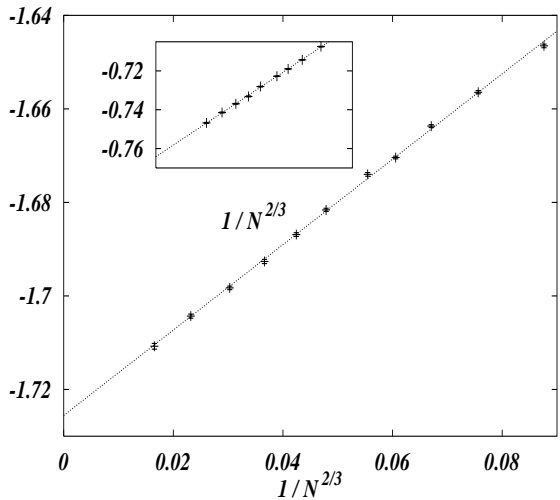}
\caption{Mean ground state energy density versus
$N^{-2/3}$ for the 
diluted mean field model with connectivity $6$ 
and for the SK model (inset). Error bars are included.}
\label{fig:Theta_s_mf}
\end{figure}

Consider first the finite
size effects in the SK model. Parisi et al.~\cite{ParisiRitort93b}
have computed analytically how various quantities
converge to their large $N$ limit. An $N^{-2/3}$ convergence
is the general rule, though for the energy density they
were able to compute the finite size correction only 
at the critical temperature $T_c$
and on the de Almeida-Thouless line. Nevertheless, the
natural extrapolation is that this law should
apply to all $T \le T_c$, leading to the prediction
$\Theta_s/d=1/3$.
(Note that this prediction is very different from that
of~\cite{CrisantiPaladin92} for which
$\Theta_s/d=\Theta_f/d=1/6$.)
To our knowledge, the possibility
that $\Theta_s/d=1/3$ for the SK ground state energy
was first brought up by Palassini~\cite{PalassiniThesis}
in his numerical studies.
We can extend his analysis with our data; 
identifying as before $L$ with $N^{1/d}$,
we perform fits of the mean ground state 
energy to Eq.~\ref{eq:Theta_s}. We then find $\Theta_s/d = 0.34 \pm 0.02$,
in complete compatibility with $1/3$; note that when plotted as
a function of $N^{-2/3}$, the data is very linear
starting from $N=50$ (see 
the inset of Fig.~\ref{fig:Theta_s_mf}).
In addition, if we perform the fit
while forcing $\Theta_s/d = 1/3$, we find that the
ground state energy density at $N=\infty$ is
$e_0=-0.7637 \pm 0.0002$, in very good agreement with 
the exact~\cite{MezardParisi87b} result $e_0 = -0.7633\ldots$

In contrast to what happens for $\Theta_f/d$,
$\Theta_s/d$ is the {\em same} in the SK model and in the 
fixed connectivity models we have considered; this
is illustrated in Fig.~\ref{fig:Theta_s_mf}.
For instance, for connectivity $6$, a power law fit
gives $\Theta_s/d=0.35 \pm 0.03$ with
$\chi^2=9.7$ for $7$ degrees of freedom.
In fact, the value $\Theta_s/d=1/3$ works very well for all four
connectivities we studied, and we are tempted to
consider that this value is the exact exponent.
The same conclusion was reached by Boettcher~\cite{Boettcher03a}.

\subsection{Edwards-Anderson models}

\begin{figure}
\includegraphics[width=8cm, height=7cm,angle=0]{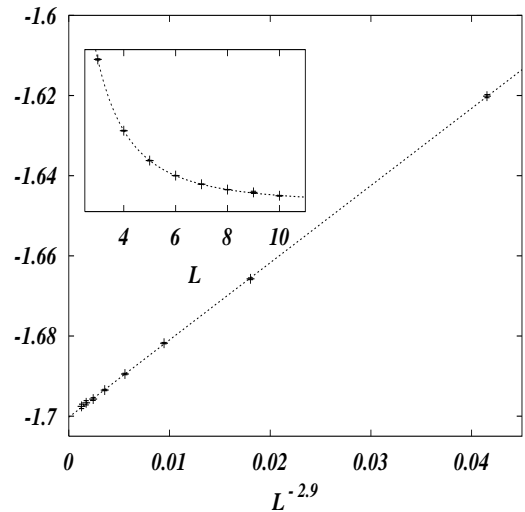}
\caption{Mean ground state energy density and the best fits for the 
$d=2$ (inset) and $d=3$ EA spin glass with Gaussian couplings.
We find $\Theta_s(d=2) \approx-0.35$ and 
$\Theta_s(d=3) \approx 0.1$. These value are ``close''
to $\theta_{DW}$, suggesting $\Theta_s=\theta_{DW}$.
Error bars are included.}
\label{fig:Theta_s}
\end{figure}

Now it is time to compare the MK and mean field
``predictions'' to our measurements 
of $\Theta_s$ in the EA models. Let us begin
with the case $d=2$. We fit our
$4 \le L \le 10$ data to Eq.~\ref{eq:Theta_s} where $e_0$,
$e_1$ and $\Theta_s$ are ajustable parameters; that gives
us the estimate $\Theta_s=-0.35 \pm 0.01$. 
(The associated $\chi^2$ is $1.9$ for $5$ degrees of freedom.)
The resulting fit is displayed along with the data
in the inset of Fig.~\ref{fig:Theta_s}.
Given our statistical error and perhaps more importantly
the systematic effects associated with $L$ not being very large, 
this value is reasonably compatible with
the MK prediction $\Theta_s=\theta_{DW}$ since
$\theta_{DW}=-0.28$~\cite{RiegerSanten96,HartmannYoung01,CarterBray02}.

A further argument in favor of $\Theta_s=\theta_{DW}$ is
as follows. It is known that whether one modifies the 
boundary conditions~\cite{PalassiniYoung99a}
or increases the size of the system~\cite{Middleton99},
the fractal dimension of the surface of the clusters of spins 
that change is approximately
the same. It is thus likely that the same type 
of excitations are involved
in determining $\Theta_s$ and $\theta_{DW}$.
As a consequence, we believe that the $d=2$ EA model
is described by a scaling theory with 
$\theta_s=\theta_{DW}$, but also with $\Theta_f \neq \theta_{DW}$
of course.
Note that the mean field prediction 
($\Theta_s = 2/3$) is clearly off; however, 
one cannot appeal to mean field when $d=2$ 
because one is below the lower critical dimension.

\begin{figure}
\includegraphics[width=8cm, height=7cm,angle=0]{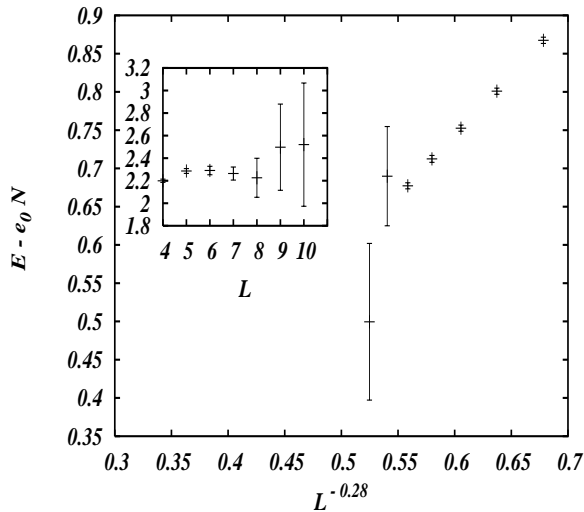}
\caption{$\overline {E_J(L)} - e_0 L^2$ 
versus $L^{-0.28}$ (main figure) and
$\overline {E_J(L)} - e_0 L^3$ versus $L$ (inset)
for the $d=2$ ($d=3$) EA 
spin glass with Gaussian couplings.}
\label{fig:Theta_s2}
\end{figure}

Let us now move on to the $d=3$ EA model which is more challenging
and has a spin glass transition at $T_c>0$. We use the same 
fitting function (Eq.~\ref{eq:Theta_s}) as before; 
the best fit then gives
a good $\chi^2$ and a mean ground state energy
growing as $-1.700 L^3 + 1.9 L^{\Theta_s}$ with
$\Theta_s=0.10 \pm 0.03$. This fit is displayed
in the main part of Fig.~\ref{fig:Theta_s}. However this value of
$\Theta_s$ easily changes by $0.1$ when removing some of the data
points, and in fact the fit sometimes even leads to negative values
for $\Theta_s$.
Thus at best we can say that $\Theta_s$ is small, somewhere
between $0.0$ and $0.2$. One can compare this result
to ``theory''. The mean field value 
($\Theta_s=1.0$) is completely
ruled out, whereas the MK prediction $\Theta_s=\theta_{DW}$ is 
quite acceptable since 
$\theta_{DW}\approx 0.2$~\cite{PalassiniYoung99b,Hartmann99}.
However, another possibility is that the discrepancy we find
has a physical origin and that in fact
$\Theta_s \ne \theta_{DW}$. Since this issue
is important, we push the analysis a bit
further as follows. Given a putative value for $\Theta_s$, we adjust
$e_0$ so that the plot of
$\overline {E_J(L)} - e_0 L^3$ versus $L^{\Theta_s}$ 
passes through the origin. For $\Theta_s$ outside
the range $\left[ 0.05,0.2 \right]$, the data has visible curvature.
In the inset of Fig.~\ref{fig:Theta_s2}
we show $\overline{E_J(L)} - e_0 L^3$ as a function of $L$;
we see that there
is no clear trend, so even $\Theta_s=0$ seems possible.
Such a value could be interpreted from the (mean-field-like)
behavior of system-size excitations found in 
this model~\cite{KrzakalaMartin00,PalassiniYoung00a}.
The analogous analysis in the $d=2$ case is shown
in the main part of Fig.~\ref{fig:Theta_s2} when we use the
value $\Theta_s=-0.28$; there the expected
corrections to scaling work quite well.

Of course it would be useful to have a significantly
smaller error bar for $\Theta_s$ 
but we cannot go much beyond what we have done
here: the statistical error on
$\overline {E_J}$ grows as $L^{d/2}$, and the amount of 
computation time grows still faster, so we cannot obtain useful
information at large $L$. 

We have no data for $d=4$, but let us briefly consider the
published work by Hartmann~\cite{Hartmann99d} where he used
the $J_{ij}=\pm 1$ EA model. (One expects its
$\Theta_s$ to be the same as in the Gaussian case.)
Analyzing his values for the mean ground state energy
for $2 \le L \le 7$, we find
$\Theta_s = 0.2 \pm 0.1$; this does not compare well
with his estimate $\theta_{DW}=0.65 \pm 0.04$. But, 
if we remove the $L=2$ point from the fit,
we find $\Theta_s = 0.7 \pm 0.2$ which is in good agreement with
$\Theta_s = \theta_{DW}$.
We have also analyzed the data of Boettcher and 
Percus~\cite{BoettcherPercus01}, and this leads
to the same conclusion.
In summary, we cannot exclude that $\Theta_s < \theta_{DW}$, 
but the MK prediction $\Theta_s=\theta_{DW}$ 
works surprizingly well
in the finite dimensional EA models. On the contrary, 
the mean field prediction
is definitely off, and that of the Mattis spin
glass is completely wrong since it gives $\Theta_s = -\infty$.

\section{Case of $+/-J$ couplings}
\label{sect:pmJ}
It is widely believed that exponents are universal, i.e., independent
of the detailed microscopic nature of the
disorder. (Note however that there are longstanding claims of universality 
violations in spin glasses~\cite{MariCampbell99}.) In particular,
for the $d=3$ EA, numerical computations of 
$\theta_{DW}$ confirm this to a large
extent: one has $\theta_{DW}=0.21 \pm 0.02$
for the Gaussian~\cite{PalassiniYoung99b} and 
$\theta_{DW}=0.19 \pm 0.02$ for the $J_{ij}=\pm 1$~\cite{Hartmann99}
models. However, if $T_c=0$ as arises in $d=2$,
one may expect several universality classes and
thus some influence of the microscopic properties 
(i.e., the distribution of the $J_{ij}$) on the
macroscopic properties (e.g., exponents). Note
first that $\theta_{DW}$ is known to be different for Gaussian and 
binary ($J_{ij}=\pm 1$) couplings~\cite{HartmannYoung01}
in the $d=2$ EA model.
{\em A posteriori}, that is not so surprising
since the associated quantities (for instance domain wall energies)
go to zero rather than to infinity; they can thus easily
be affected by microscopic details. 
This issue can be investigated
within the framework of MK lattices. We find
identical values of $\theta_{DW}$ in the Gaussian 
and the $J=\pm 1$ cases whenever 
$\theta_{DW} > 0$. However in $d=2$ and using
$b=\ell=2$, the Gaussian model gives 
$\theta_{DW} = - 0.22$; on the contrary,
the $J_{ij}=\pm 1$ case leads to 
$\theta_{DW}=-\infty$, meaning that the domain-wall 
energies decrease exponentially
with the size of the system rather than as a power law.
Thus either there is violation of universality
(which seems unlikely to us) or there are several
universality classes when $\theta_{DW} \le 0$;
Amoruso et al.~\cite{AmorusoMarinari03} 
have given evidence in favor of the latter
possibility since the first version of this paper was posted.

\begin{figure}
\includegraphics[width=8cm, height=7cm,angle=0]{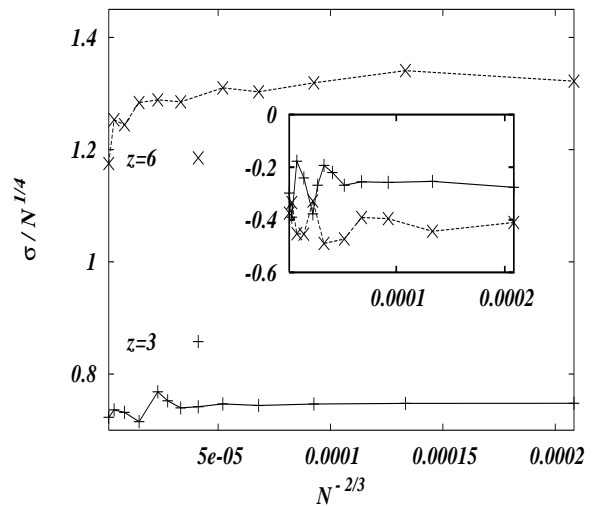}
\caption{Standard deviation of $E_J$ divided by 
$N^{1/4}$ for the $z=3$ and $6$ fixed connectivity
models, but with binary couplings, $J_{ij}=\pm 1$.
Inset: skewness of the distribution of the ground
state energy for these same two models.}
\label{fig:sigmaFCPMJ}
\end{figure}

Given these amendments to the scope of universality, 
the microscopic details should indeed
be irrelevant for quantities associated 
with {\em diverging} energy scales. The surprising 
claim we bring forward here
is that this expectation is still too strong: a 
counter example is provided by 
$\Theta_f$ in the mean field fixed connectivity graphs.
Indeed, we saw in the case of 
Gaussian couplings that $\Theta_f/d=1/2$. 
Now in Fig.~\ref{fig:sigmaFCPMJ} we show the
rescaled standard deviation of $E_J$ for the case where
$J_{ij}=\pm 1$ on fixed connectivity graphs; the
standard deviations have been divided by $N^{1/4}$. If
$\Theta_f/d = 1/2$, we should see a rapid divergence
of the plotted values with increasing $N$,
but instead the curves are relatively flat and decreasing!
In fact, when performing fits, we find that
$\Theta_f/d$ is between $1/4$ and $1/5$.

To give further evidence that the 
Gaussian and $J_{ij}=\pm 1$ cases scale differently,
we plot in the inset of this figure the values of the skewness
of the distribution of $E_J$ using the same models
and values of $N$ as in the main figure. Although
our data is noisy, the skewness shows no
sign of going to zero when $N \to \infty$.
Taken at face value, this means that the
distribution of $E_J$ is not Gaussian in
the large $N$ limit when $J_{ij}=\pm 1$, in sharp 
contrast to what happens 
in the case of Gaussian couplings.
To drive this point home further, we find that
as $z$ increases,
the skewness and kurtosis grow in magnitude and
seem to approach the values we find in the SK. 
(But as before, our values become imprecise at large $N$.)
We thus conjecture
that each $z$ is associated with a different universality
class and that, as $z \to \infty$, one converges to
the class to which the SK model belongs. One way to justify this
is to consider that in these models $z$ is related to dimension
rather than to lattice connectivity;
the universality class will then change with $z$.
To a large extent, all different trends give further
credence to the claim that $\Theta_f/d = 1/4$ in the SK model
if that is the correct value for the fixed connectivity models
with binary couplings.

How can one understand this ``break-down'' of universality
when the underlying energies diverge?
Recall that for the $J_{ij}=\pm 1$ spin
glass model on a fixed connectivity graph,
the local environment of a spin has no disorder out to
finite distances: any sign of the $J_{ij}$ can be gauged away so that
all the sample-dependent fluctuations arise ``at infinity''. On the contrary, 
in the Gaussian case the $J_{ij}$ fluctuations are local, leading to
$O(N^{1/2})$ fluctuations in the total energy. 
Taken at face value, this argument also applies to
{\em coordination} fluctuations; if this is true, the $\pm 1$
Viana-Bray model~\cite{VianaBray85} will have
$\Theta_f=d/2$.

To summarize, the exponent $\Theta_f$ depends on the
details of the underlying $J_{ij}$ 
distribution even though the energy scale (of fluctuations)
diverges. Not surprisingly, we also find that
$P_N(E_J)$ for that system does not become Gaussian.
Finally, in spite of this major change of behavior when going
from Gaussian $J_{ij}$ to binary values,
we find that $\Theta_s/d = 1/3$ very precisely in both cases.

\section{Discussion and conclusion}
\label{sect:discussion}

Let us go over the main results of the present work
on spin glasses. First, the sample to 
sample fluctuations of the ground state energy
grow as the square-root of the volume in almost
all models so the fluctuation
exponent satisfies $\Theta_f=d/2$. Furthermore, the 
distribution over disorder of the
ground state energy probably
tends towards a Gaussian in the large volume limit as suggested by
Wehr and Aizenman~\cite{WehrAizenman90}. There are two
notable exceptions to this picture: the SK model and
the fixed connectivity 
mean field models having $J_{ij}=\pm 1$. In those
two cases, the sample to sample fluctuations are much smaller, 
and we find $\Theta_f\approx d/4$, even though
a still smaller value cannot be excluded.
(Note that fluctuations which are smaller than square root
of the volume $N$ are also a characteristic
of the directed polymer in a random medium.)
On the other hand, for finite dimensional lattices, one expects 
the finite density of unfrustrated regions
to contribute $O(N^{1/2})$ fluctuations to the ground state energy.
A trivial example where this is the case is the (unfrustrated) Mattis model;
there the variance of the ground state energy is obviously maximal and 
equal to $z N \left[ \langle J^2 \rangle -\langle J \rangle^2 \right]/2$
for a connectivity $z$.

We also studied the exponent $\Theta_s$ giving the corrections to 
the scaling of the average (extensive) ground state energy. 
For the $d=2$ and $3$ EA models, we find that
the equality $\Theta_s=\theta_{DW}$ holds within
our limited precision. This means that corrections to scaling
are associated with domain walls.
However, we were not able to rule out $\Theta_s=0$ in $d=3$, leaving
the door open to other interpretations.
In the context of the MK lattices, the equality 
$\Theta_s=\theta_{DW}$ holds exactly.
Physically, this equality corresponds to the fact that in 
a finite fraction of the samples, the boundary conditions force a domain 
wall ``defect'' to be present in the ground 
state. (Note that the finite size corrections are indeed always positive
for fixed and periodic boundary conditions). Furthermore, we were able to 
compute $\theta_{DW}$ analytically to order $1/b$ when the dimension
becomes large.
In the case of the 
mean field models (SK or diluted graphs), the situation is quite different:
the corrections to scaling grow as $N^{1/3}$. Since
one expects $\theta_{DW} \sim d/2$ in
large dimensions, the result $\Theta_s/d=1/3$ for these mean
field models differs from
the large dimension limit of $\theta_{DW}/d$.

Finally, we have exhibited examples where the 
exponent $\Theta_f$ depends
on the distribution of the $J_{ij}$, even though energy fluctuations
diverge with size, i.e., $\Theta_f>0$.

Our work suggests several
paths for further studies.
(a) Can one establish the value of $\Theta_f/d$ for the SK 
model? (b) Since large fluctuations of order $N^{1/2}$ are 
detrimental in a numerical determination of 
the average ground state energy, is there a way to 
substract a (computable) contribution
of satisfied bonds so as to reduce the variance?
(c) Is $\Theta_s=\theta_{DW}$ in finite dimensional
spin glasses or is $\Theta_s$ smaller? This second
possibility could follow from other 
types of excitations whose exponents are smaller than $\theta_{DW}$.

\paragraph*{Acknowledgments ---}

We thank A. Billoire, K. Binder, D. S. Fisher, 
M. M\'ezard, M. Palassini, G. Parisi, R. da Silveira 
and O. White for stimulating discussions.
This work was supported in part by the European Community's
Human Potential Programme under the 
contracts HPRN-CT-2002-00307 and 
HPRN-CT-2002-00319.
F.K. acknowledges financial support from the MENRT.
The LPTMS is an Unit\'e de Recherche
de l'Universit\'e Paris~XI associ\'ee au CNRS.

\section*{Appendix: Exponents in the Migdal-Kadanoff lattices
and the large dimensional limit}

In this appendix, we focus on the simplest case $\ell=2$. To 
write a recursion
relation, one needs to keep track of two energies, $E_p^{(r)}$ 
and $E_a^{(r)}$ that give the
ground state energy of the MK lattice at the $r$-th application
of the recursion when the 
two exterior spins 
are respectively parallel and antiparallel. For a $b$-branch 
lattice with $\ell=2$, 
the ground state energies at the $(r+1)$-th application of the recursion read:
\begin{eqnarray}
E_p^{(r+1)} = \sum_{\alpha=1}^b \min \nonumber \\
\left(E_p^{(r)}(1,\alpha)+E_p^{(r)}(2,\alpha),E_a^{(r)}(1,\alpha)+E_a^{(r)}(2,\alpha)\right)\\
E_a^{(r+1)} = \sum_{\alpha=1}^b \min \nonumber \\
\left(E_p^{(r)}(1,\alpha)+E_a^{(r)}(2,\alpha),
E_a^{(r)}(1,\alpha)+E_p^{(r)}(2,\alpha)\right)
\end{eqnarray}
where the index $1,2$ refers to the two bonds in the $\ell$ direction. This 
equation says that along all the $b$ branches, one has to choose the 
orientation of the
middle spin (that is decimated) such that the energy contribution 
is minimized, given the orientations of the external spins. Note 
that all the random variables that
appear in this equation are independent as soon as they live 
on different bonds.

If one assumes that the distribution $P^{(r)}(E_p,E_a)$
of energies at the $r$-th 
generation can be written for large $r$ in the scaling 
form 
\begin{equation}
P^{(r)}(E_p,E_a)=1/\sigma_r^2 f(\left[ E_p-{\cal E}_r \right]/\sigma_r, 
\left[ E_a-{\cal E}_r \right] /\sigma_r )
\end{equation}
then it is immediate to show (using the independence of 
the branches) that ${\cal E}_{r+1}=(2b) {\cal E}_r$ and 
$\sigma_{r+1}=\sqrt{2b}\sigma_r$, so 
that ${\cal E}_r \sim (2b)^r$ and $\sigma_r \sim (2b)^{r/2}$. Since 
the number of spins is given by $(2b)^r$, 
one immediately finds that the ground state energy is extensive and 
that the fluctuations are described by $\Theta_f/d=1/2$. 

Define $\Delta^{(r)}$ as $E_p^{(r)}-E_a^{(r)}$. It is easy to
show that $\Delta^{(r)}$ obeys an autonomous recursion 
relation~\cite{SouthernYoung77}:
\begin{equation}
\Delta^{(r+1)} = \sum_{\alpha=1}^b \epsilon(\alpha) 
\min\left(|\Delta^{(r)}(1,\alpha)|,|\Delta^{(r)}(2,\alpha)|\right),
\end{equation}
where $\alpha$ labels branches and
\begin{equation}
\epsilon(\alpha) = - {\rm{sign}}\left[\Delta(1,\alpha)\right] \ 
{\rm{sign}}\left[\Delta(2,\alpha)\right] \ .
\end{equation}
Using the independence of the $\Delta$'s 
corresponding to different branches, one finds that $\Sigma_r$,
the RMS of the distribution of the
$\Delta$'s, obeys the following equation:
\begin{equation}
\Sigma_{r+1}^2 =  4 b \Sigma_{r}^2 \int_0^\infty \, {\mbox d}x g(x) \left[
\int_0^x \, {\mbox d}y y^2 g(y) +  x^2 \int_x^\infty \, {\mbox d}y g(y)\right],
\end{equation}
where $g(.)$ is the distribution of $\Delta/\Sigma$ that 
is independent of $r$ for
large $r$. From this relation, one finds 
that $\Sigma_r \sim \lambda^r$ with
\begin{equation}
\label{eq:lambda}
\lambda^2 =  4 b \int_0^\infty \, {\mbox d}x g(x) \left[
\int_0^x \, {\mbox d}y y^2 g(y) +  x^2 \int_x^\infty \, {\mbox d}y g(y)\right]
\end{equation}
where of course $g(.)$ depends on $b$. The energy scale for
flipping the relative sign 
of the exterior spins is $\Sigma$ and 
so the exponent $\theta_{DW}$
is given by
\begin{equation}
\theta_{DW}= \frac{\ln \lambda}{\ln 2} \ .
\end{equation}
In the large dimension limit,
for which $b \to \infty$, it is clear that, using the central limit theorem, 
the distribution of $\Delta$ is Gaussian. Since
$g(.)$ is then known, $\lambda$ can 
be computed from Eq.~\ref{eq:lambda}, giving
\begin{equation}
\lambda^2 =  0.36338~b,
\end{equation}
a result first obtained by Southern and Young.

When $b$ is large but not infinite, the first 
correction to the Gaussian is of order $1/b$ 
(because the $\Delta$ have a symmetric distribution) and reads:
\begin{equation}
g(x)= \frac{1}{\sqrt{2\pi}} 
\left[1+ \frac{\kappa}{24 b} \frac{\partial^4}{\partial x^4} + ...\right]
e^{-x^2 / 2}
\end{equation}
where $\kappa$ is the kurtosis of the initial variable, {\em i.e.}, $\epsilon
\min\left(|\Delta(1)|,|\Delta(2)|\right)$. To first order 
in $1/b$, this kurtosis can 
be computed by assuming that the $\Delta$ are Gaussian, and 
one finds $\kappa=0.434215$.
Injecting the expression of $g(x)$ in Eq.~(\ref{eq:lambda}) 
then gives $\lambda$ to order
$1/b$: $\lambda^2 =  0.36338~b - 0.023035$, and finally the 
result given in the main text after 
Eq.~(\ref{eq:MKresult}).

This calculation can be extended to next order: $\kappa$ will 
acquire a $1/b$ contribution
and there will be corrections to $g(x)$ of
order $1/b^2$ coming from the non zero sixth
cumulant of $\epsilon \min\left(|\Delta(1)|,|\Delta(2)|\right)$, computed 
as if the 
$\Delta$'s were Gaussian. One could also, with more work, 
compute $\lambda$ for $\ell \neq 2$.
In the limit $\ell \to \infty$, the problem becomes soluble 
again using the theory of 
extreme value statistics for handling the 
variable $\min\left(|\Delta(1)|,|\Delta(2)|, ..., |\Delta(\ell)|\right)$.

\bibliographystyle{prsty}
\bibliography{../../../Bib/references}

\end{document}